\documentclass[a4paper,12pt]{article}

\usepackage{amsmath}

\usepackage{amssymb}

\usepackage{makeidx}

\begin{document}

\newcommand{\intx}{\int d^4 x\,}

\newcommand{\inty}{\int d^4 y\,}

\newcommand{\intk}{\int d^4 k\,}

\newcommand{\intp}{\int d^4 p\,}

\newcommand{\meio}{\frac{1}{2}}

\newcommand{\tr}{\,\text{tr}\,}

\begin{titlepage}

\begin{center}   
{\Large\bf
Comment on ``A new, exact, gauge-invariant RG-flow equation''}  
\end{center} 
\vspace{0.5cm} 
\begin{center} 
{\large Filipe \hspace{-0.07cm}Paccetti \hspace{-0.07cm}Correia\footnote{E-mail address: 
F.Paccetti@ThPhys.Uni-Heidelberg.DE}  
} 
\vspace{0.5cm}

{\em Institut f\"ur Theoretische Physik, Universit\"at Heidelberg,
Philosophenweg 16,\\
 D-69120 Heidelberg, Germany \\

} 

\end{center}

\vspace{1.0cm} 

\begin{abstract}
We show that the exact RG-flow equation introduced re\-cent\-ly in hep-th/ 0207134 can be obtained in the sharp cut-off limit of the well-known ERGE. This can be expected from the fact that in this limit the new scale-dependent effective action coincides with the one which is usually considered.  
\end{abstract}

\end{titlepage}

The purpose of this short note is to show that the \emph{new} exact RG-flow equation recently proposed in ref.\cite{Bra02}, which is obtained by a partial Legendre transform , corresponds to the \emph{sharp} cut-off limit of the \emph{exact} RG equation (ERGE) of ref.\cite{wet93,bon93}~\footnote{~For a comprehensive review see \cite{ber02}}. To do this we will start by deriving the ERGE before considering the sharp cut-off limit.

\paragraph{The effective action.} The scale-dependent 1PI effective action can be defined in the following way: One introduces the quadratic cut-off functional
\begin{equation}
                 {\mathcal O}_k [\chi-\varphi]\equiv \exp{\left(-\Delta_k S[\chi-\varphi]\right)}\equiv\exp{\left( -\meio\int(\chi-\varphi)R_k(\chi-\varphi)\right)},
\end{equation}
inside the functional integral which defines the partition function,
\begin{equation}
                 e^{W[J]}\equiv\int{\mathcal D}\chi\, e^{-S[\chi]+\int J\chi}.
\end{equation}
We assume that $\lim_{k\to 0}R_k=0$ and therefore the functional
\begin{equation}
                 e^{\widehat{W}_k [J,\varphi]}\equiv\int{\mathcal D}\chi\, {\mathcal O}_k[\chi-\varphi]\, e^{-S[\chi]+\int J\chi},
\end{equation}
obtained with this procedure converges to $W[J]$ as $k\to 0$. We can now perform a Legendre transformation of $W_k$ with respect to $J$ to obtain
\begin{equation}
                 \widehat{\Gamma}_k[\phi,\varphi]=-\widehat{W}_k[J,\varphi]+\int J \phi, \quad \phi \equiv \frac{\delta \widehat{W}_k[J,\varphi]}{\delta J}. 
\end{equation}
This means that $\widehat{\Gamma}_k$ is given implicitly by
\begin{equation}\label{impl}
                \exp{\left(-\widehat{\Gamma}_k[\phi,\varphi]\right)} = \int{\mathcal D} \chi \, {\mathcal O}_k [\chi-\varphi+\phi] \, \exp{\left(-S[\chi+\phi]+\int\frac{\delta \widehat{\Gamma}_k [\phi,\varphi]}{\delta \phi}\chi\right)}.
\end{equation}

Since it is our intention to obtain an effective action which interpolates between the classical action $S[\phi]$ in the $UV$ ($k\to\infty$), and the full effective action $\Gamma[\phi]$ in the IR ($k=0$) we first assign ${\mathcal O}_k[\chi-\varphi]$ the following property
\begin{equation}
                  \lim_{k\to\infty} {\mathcal O}_k[\chi] \sim \delta [\chi],
\end{equation}          
where the r.h.s. is a $\delta$-functional. This can be obtained if $\lim_{k\to\infty} R_k =\infty$. In this case we have
\begin{equation}
                  \lim_{k\to\infty} \widehat{\Gamma}_k[\phi,\varphi] = S[\varphi] + \int\frac{\delta \widehat{\Gamma}_k[\phi,\varphi]}{\delta \phi}(\phi-\varphi).
\end{equation}
Finally, to obtain the desired property, we set $\varphi=\phi$:
\begin{equation} 
                  \Gamma_k[\phi]\equiv\widehat{\Gamma}_k[\phi,\phi].
\end{equation}
One can use now
\begin{equation}\label{deriv}
           \frac{\delta \widehat{\Gamma}_k[\phi,\varphi]}{\delta\varphi}=-\frac{\delta \widehat{W}_k[J,\varphi]}{\delta\varphi}=R_k(\varphi-\phi),
\end{equation}
to write $\Gamma_k[\phi]$ as (see eq.\eqref{impl})
\begin{equation}
                \exp{\left(-\Gamma_k[\phi]\right)}=\int{\mathcal D}\chi\,{\mathcal O}_k[\chi]\,\exp{\left(-S[\chi+\phi]+\int\frac{\delta \Gamma_k[\phi]}{\delta \phi}\chi\right)}.
\end{equation}
From this expression it is not dificult to recognize $\Gamma_k[\phi]$ as being given (perturbatively) by the sum of the connected 1PI vacuum graphs with $\phi$-dependent vertices and internal lines regulated by the introduced cut-off. 

The attentive reader may consider the above formalism reminiscent of the background field method used to ensure the gauge invariance of the 1PI effective action (see~\cite{abb81} and refences therein). And indeed both ideas can be combined to define a scale-dependent, gauge invariant, 1PI effective action~\cite{reu94}.

\paragraph{The flow equation.} To calculate the exact renormalization group equation (ERGE) for $\Gamma_k$ let us note that
\begin{equation}\begin{split}
                 \partial_k \Gamma_k[\phi] & =\partial_k \widehat{\Gamma}_k[\phi,\phi] = - \partial_k \widehat{W}_k [J,\phi] =\meio\int\partial_k R_k \langle (\chi-\phi)(\chi-\phi)\rangle\\
          & = \meio\int\partial_k R_k\frac{\delta^2 \widehat{W}_k[J,\phi]}{\delta J\delta J}.
\end{split}\end{equation}
Now, we have
\begin{equation}
             \frac{\delta^2 \widehat{W}_k[J,\phi]}{\delta J\delta J}=\left( \frac{\delta^2 \widehat{\Gamma}_k[\phi,\varphi]}{\delta\phi\delta\phi} {\Big |}_{\varphi=\phi}\right)^{-1},
\end{equation}
while
\begin{equation}
           \frac{\delta^2 \Gamma_k[\phi]}{\delta\phi\delta\phi}= \frac{\delta^2 \widehat{\Gamma}_k[\phi,\varphi]}{\delta\phi\delta\phi}{\Big |}_{\varphi=\phi} + 2 \frac{\delta^2 \widehat{\Gamma}_k[\phi,\varphi]}{\delta\phi\delta\varphi}{\Big |}_{\varphi=\phi} + \frac{\delta^2 \widehat{\Gamma}_k[\phi,\varphi]}{\delta\varphi\delta\varphi}{\Big |}_{\varphi=\phi}.
\end{equation}
Using eq.\eqref{deriv} one gets
\begin{equation}
           \frac{\delta^2 \widehat{\Gamma}_k[\phi,\varphi]}{\delta\phi\delta\phi}{\Big |}_{\varphi=\phi}= \frac{\delta^2 \Gamma_k[\phi]}{\delta\phi\delta\phi} + R_k,
\end{equation}
obtaining in this way the well-known ERGE for $\Gamma_k$~\cite{wet93,bon93}:
\begin{equation}\label{floweq}
           \partial_k \Gamma_k[\phi] = \meio\int(\partial_k R_k)\frac{1}{\Gamma_k^{(2)}[\phi] + R_k}.
\end{equation}

\paragraph{Sharp cut-off limit.} By sharp cut-off one means a function $R_k$ which diverges for momenta below the scale $k$ and vanishes above this scale. Although one could think that in this limiting case $\Gamma_k$ is not well defined due to an ill-definition of the Legendre transform of $W_k[J,\varphi]$ we will see that this is not the case. 

In this limit ${\mathcal O}_k$ is of the form
\begin{equation}
               {\mathcal O}_k [\chi] \sim \prod_{p^2<k^2} \delta(\chi(p)),
\end{equation}
which means that (using the notation of ref.\cite{Bra02}) we can write
\begin{equation}
               \exp{W_k[J,\varphi_0^k]}=\int{\mathcal D}\chi_k^{\Lambda}\,\exp{\left(-S[\varphi_0^k+\chi_k^{\Lambda}]+\int J(\varphi_0^k+\chi_k^{\Lambda})\right)},
\end{equation}
where $\varphi_0^k$ contains only Fourier modes with $p<k$ and $\chi_k^{\Lambda}$ only modes with $p>k$. This differs from the $W_k$ ($\equiv {\mathcal W}_k$) defined in~\cite{Bra02} in the following way\footnote{~There is also a irrelevant difference in sign.}
\begin{equation}
               W_k [J,\varphi_0^k]=  {\mathcal W}_k [J_k^{\Lambda},\varphi_0^k] +\int J_0^k \varphi_0^k,
\end{equation}          
but, as we will show, the scale dependent effective action $\Gamma_k$ coincides with the one defined in that work (we will call this one ${\mathcal G}_k$). This follows from the fact that $\Gamma_k[\phi]= \widehat{\Gamma}_k[\phi,\phi]$ in this case is given by
\begin{equation}
               \Gamma_k[\phi]=\widehat{\Gamma}_k[\phi,\varphi_0^k],
\end{equation}
since due to the sharp cut-off we have $\phi=\varphi_0^k$ for $p<k$ and $W_k$ is independent of $\varphi_k^{\Lambda}$, where $ \varphi_k^{\Lambda}=\varphi$ for $p>k$. Furthermore, we have
\begin{equation}
               \Gamma_k[\phi]=-W_k[J,\varphi_0^k]+\int J(\varphi_0^k+\phi_k^{\Lambda})=-{\mathcal W}_k [J_k^{\Lambda},\varphi_0^k]+\int J_k^{\Lambda}\phi_k^{\Lambda},
\end{equation}
where
\begin{equation}
               \phi_k^{\Lambda}=\frac{\delta {\mathcal W}_k [J_k^{\Lambda},\varphi_0^k]}{\delta J_k^{\Lambda}}.
\end{equation}     
But this is clearly the definition of ${\mathcal G}_k$, the Legendre transform of ${\mathcal W}_k$ with respect to $J_k^{\Lambda}$, as we intended to show.

\paragraph{The flow equation in the sharp cut-off limit.} Let us now see what happens to the flow equation \eqref{floweq} in the sharp cut-off limit. The flow equation can be rewritten as
\begin{equation}
               \partial_k \Gamma_k[\phi] = \meio\tr \tilde{\partial}_k \ln{(\Gamma_k^{(2)}[\phi] + R_k)}=\meio\tilde{\partial}_k\ln\det{(\Gamma_k^{(2)}[\phi] + R_k)},
\end{equation}
where $\tilde{\partial}_k$ only acts upon $R_k$. Since below $k$ the cut-off fuction $R_k$ diverges we get
\begin{equation}
               \partial_k \Gamma_k[\phi] = \lim_{\delta k\to 0}\frac{1}{2\delta k} \left[\ln\det\frac{\delta^2 \Gamma_k}{\delta \phi^{\Lambda}_{k}\delta \phi^{\Lambda}_{k}}-\ln\det\frac{\delta^2 \Gamma_k}{\delta \phi^{\Lambda}_{k-\delta k}\delta \phi^{\Lambda}_{k-\delta k}}\right],
\end{equation}
where the $k$ in the first term and the $k-\delta k$ in the second one denote the IR cut-offs in the respective determinants and we neglected a $\phi$-independent piece. It is not dificult to recognize that this is the flow equation which was obtained by the authors of ref.\cite{Bra02} by other means.\\

Thus in this paper we showed that the scale-dependent 1PI effective action defined in \cite{Bra02} by a partial Legendre transformation can be considered to be the sharp cut-off limit of the one defined in \cite{wet93,bon93}. As expected, the flow equation of \cite{Bra02} could also be obtained as the same limit of the ERGE of \cite{wet93,bon93}.\\

{\bf Acknowledgements} 

The author wishes to thank T.Baier, M.G.Schmidt and Z.Tavartkiladze for usefull discussions. This work is supported by Funda\c c\~ ao de Ci\^ encia e Tecnologia (grant SFRH/BD/4973/2001).

\end{document}